\documentclass[10.5pt]{article}

\usepackage{mathptmx}
\usepackage{graphicx}
\usepackage{times}
\usepackage{amsmath, amsthm, amssymb}
\usepackage{url}
\usepackage{subfig}
\usepackage{hyperref}
\usepackage{algorithm}
\usepackage{algorithmic}

\usepackage{lscape}

\begin{document}

\begin{titlepage}

\begin{center}
{ \Large \bf
{Agent-based simulations of emotion spreading in online social networks}
}

{\large Milovan \v Suvakov$^{1,2}$,  David Garcia$^{3}$,  Frank Schweitzer$^3$ and Bosiljka Tadi\'c$^1$}\\
{ $^1$Department of theoretical physics; Jo\v zef Stefan Institute;
 Box 3000  SI-1001 Ljubljana Slovenia; $^2$Institute of Physics, Belgrade, Serbia,  $^3$Chair of Systems Design, ETH Z\"urich, Switzerland\hspace{1cm}}

\date{today}

\end{center}

\noindent
\begin{abstract}
Quantitative analysis of  empirical data from online social networks reveals group dynamics in which emotions are involved \cite{we-MySpace11}. Full understanding of the underlying mechanisms, however, remains a challenging task.
Using agent-based computer simulations, in this paper we study dynamics of emotional communications in online social networks. The rules that guide how the agents interact are motivated, and the realistic network structure and some important parameters are inferred from the empirical dataset of \texttt{MySpace} social network.  
Agent's emotional state is characterized by two variables representing psychological arousal---reactivity to stimuli, and valence---attractiveness or aversiveness, by which common emotions can be defined.  Agent's action is triggered by increased arousal. High-resolution dynamics is implemented where each message carrying agent's emotion along the network link is identified and its effect on the recipient agent is considered as continuously aging in time. 
Our results  demonstrate that (i) aggregated group behaviors may arise from individual emotional actions of agents; (ii)  collective states  characterized by temporal correlations and dominant positive emotions emerge, similar to the empirical system; (iii) nature of the driving signal---rate of user's stepping into online  world, has profound effects on building the coherent  behaviors, which are observed for users in online social networks. Further, our simulations suggest that spreading patterns differ for the emotions, e.g., ``enthusiastic'' and ``ashamed'', which have entirely different emotional content.
{\bf {All data used in this study are fully anonymized.}}

\end{abstract}
\vspace{0.5cm}
\tableofcontents
\end{titlepage}
\newpage

\section{Introduction\label{sec-intro}}
The article ``Postmodernity and affect: All dressed up with no place to
go'' published by Lawrence Grossberg in 1988 addresses the altered view
of emotions in the postmodern society \cite{grossberg}. But ``the place
to go'' has considerably changed since. Instead of meeting other people
at preferred locations, the Internet has become the most important
``place'' to show oneself dressed up, either as a real-name person or
with a fake identity, but still with certain desires, goals, or feelings. 
In fact, the Internet is increasingly recognized not only as a tool that
people use, but as an environment where they function and live.  The
amount of time, energy and emotions spent on using  online
social networks  \cite{we-MySpace11,giles2010,amichai-hamburger2010,cheung2011,ryan2011}, online games  \cite{szell2010,ST12}, emails \cite{albert_emailnets}, blogs \cite{mitrovic2010b} and chats \cite{garas2011} are getting unprecedented scores. 
Hence the social implications of the Internet  \cite{DiMaggio2001,amichai-hamburger2002,johnson2009}, the intricate relationships between the real and the virtual worlds \cite{szell2010,ST12},  and mechanisms governing new techno-social phenomena \cite{kleinberg2008,panzarasa2009,dodds2011,mitrovic2010a,mitrovic2010c}  pose new challenges for scientific research.

 The social sciences   faced with the problem of transferring concepts of ``offline'' human behaviors into the online world of social networks \cite{nature2011}. The question whether humans behave completely different when becoming ``users'' is tackled in various empirical investigations \cite{DiMaggio2001,Sassenberg2003,johnson2009,garcia2011,garas2011,mitrovic2011}. A particular question regards the emotional interaction between users in online social networks: How is emotional influence exerted if mostly written text is exchanged? What kind of emotions are actually involved? What is the role of the underlying network structure in spreading emotions? 
Therefore, study of the stochastic processes  related with stepping of the users into the virtual world and spreading of their behaviors and emotions through the online social network, are  of key importance. 
 
In our recent work \cite{we-MySpace11}, hereafter referred as I,  we have compiled and analysed large dataset containing the dialogs from  \texttt{MySpace} social network, currently ranked as the fourth largest social networking site after \texttt{Facebook}, \texttt{Google+}, and \texttt{Linkedin}.
Combining the methods of statistical physics with machine learning approach of text analysis, by which  the emotional content of the messages was extracted, we have found strong evidence of user's collective behaviors in which emotions are involved. Specifically, bursts of emotional messages occur, which obey scaling laws and temporal correlations. Furthermore, dominance of positive emotions and specific structure of the dialogs-based network was revealed. 
In this work we use agent-based framework with emotional agents \cite{schweitzer2010}, and explicitly investigate the mechanisms of the emotional influence between agents in the online social  network, which is underlying the observed collective behaviors of users.

In general, we observe that many models of social interaction are based on the KISS approach (``Keep it simple and stupid''). 
In contrast to these simple approaches, our model is specifically designed to
describe the emotional interaction in \texttt{MySpace}, which is
certainly different from other online platforms, such as blogs, or
fora. We do not try to simplify the system and this way get rid of all
the specific features. On the contrary, we want to build up a model that
as much as possible includes these specific features, such as,
e.g., directed communication through message walls. Consequently, some of
the assumptions used in our model read quite complicated on the first
glimpse, but try to capture real interactions in larger detail. This
restricts of course the possible generalization to other online platforms,
which is not the aim of the paper.

Agent based approaches are gaining importance among the psychologist community
\cite{Rodgers2010, Kuppens2010}. Our modeling framework has already
proved its applicability for various online communications,
e.g. emotional influence in chats \cite{garas2011}, product reviews
\cite{schweitzer2010,garcia2011b} and dynamics of blogs \cite{mitrovic2011}. It is based on the concept of Brownian agents
\cite{schweitzer-book} which are described by two scalar variables,
\emph{valence} that describes the pleasure ( attractiveness and aversiveness) associated with an emotion,  and \emph{arousal} that describes the activity level induced by the emotion. Our focus on valence and arousal to quantify emotions is
motivated by Russel's model \cite{russell1980, scherer2005,emotions-Oxford2007} from psychology.  In our modeling framework
stochastic equations for both variables are proposed to account for
random influences. Their deterministic parts result from the direct and
indirect emotional interaction between agents and needs to be specified
according to the online system considered. Importantly, the nonlinear
functions of the deterministic part capture the response of an agent to
emotional information. The latter is modeled as a time-dependent field
that is generated by agents posting emotional text, this way comprising
valence and arousal of different agents.  The response functions specify
what of this information field actually affects a specific agent. While
in some cases, such as fora communication, a mean-field approach is most
appropriate, i.e. all agents can see all information, the new feature of
the model proposed here is in the network communication, i.e. agents can
only see very specific information of ``friends'', which are different
for every agent. This makes the model more complicated compared to fora
communication, but it still follows the general outline of the framework
of emotional influence. 
In this respect, mathematical complexity of the present model, with the \textit{emotional agents on a fixed network},  lies in between two previously studied cases. On one side, the dynamics of Blogs where the  agent's actions cause the network on which they are situated to evolve \cite{MMBT_ABM2011,MMBT_abmnetworks}, and on the other, the product review communications where the  agents are not exposed to any geometrical constraints  \cite{schweitzer2010}. 

The model proposed in this paper is linked to empirical observations in
\texttt{MySpace} in various ways. First, it takes some empirical findings
as \emph{input} for the computer simulations, in particular the network of
interactions which has been extracted from the dataset
\cite{we-MySpace11}. To have this as a realistic feature is quite
important because hidden topology features, such as link correlations at
next-neighborhood level, affect the spreading of information and other
relaxation dynamics in complex networks
\cite{rocaPLOS2011,IJBCH2007,PhysRevLett.94.137204}. A second important
empirical input regards the temporal activity patterns of users, which
are known to display some universal features across different
communication media \cite{malmgren2009,castellano2009,crane2009,mitrovic2010a,mitrovic2010c,szell2010}. 

The \emph{output} of our computer simulations is also to be compared to
empirical findings. Importantly, the creation and exchange of all
emotional messages results in a valence distribution, which allows a
comparison of empirics and simulation on the aggregated level. For the
empirical analysis, the emotional information captured in the user's
messages has been extracted by means of sentiment analysis algorithms
\cite{thelwall2010, paltoglouIEEE2011}. 
In contrast to the analysis in Ref.\ I,  where only the emotion \emph{valence} is considered and accurately graded by the emotion classifiers \cite{thelwall2010, paltoglouIEEE2011}, for the comparison of the simulated and empirical data here we need a new classifier in which both \emph{valence and arousal} of the messages are determined simultaneously.  Therefore we introduce a lexicon-based classifier which is not as accurate as the valence classification, but it  provides two values, valence and arousal, from a written text. Note that, according to psychology literature \cite{russell1980, scherer2005,emotions-Oxford2007}, these two components can already define a commonly known emotion, e.g., ``fear'', ``shame'', ``delight'', etc. 
Secondly, our model simulates the response
of the social network to external emotional events which allows to
quantify the time scale for the emergence of collective emotions. The
resulting power spectrum can be compared to empirical data.

In the following section, we first describe the empirical findings from
the \texttt{MySpace} dataset that are relevant for the modeling
part. The agent based model and its numerical implementation are
described in Sect. \ref{sec-model}, while the results of computer
simulation and their comparison to empirical findings are presented in
Sect. \ref{sec-simulations}. A general discussion concludes the paper.

\section{Input from the empirical dataset of \texttt{MySpace}\label{sec-empirical}}

\subsection{Structure of the underlying network\label{sec-network}}
As mentioned in the Introduction, the underlying network structure may crucially affect the diffusion processes on it, e.g.,  by imposing ``topology traps'' and channeling the process according to its higher topology structures. Therefore, to obtain reliable dynamic features of the emotion spreading process, the realistic network structure is considered. 
The network we use for the modeling in Sect. \ref{sec-model} is obtained from
the social networking site \texttt{Myspace}.  The data are collected by a parametrized algorithm with a specified depth (time-window of the dialogs) and the network diameter in paper I \cite{we-MySpace11}.
In  \texttt{Myspace}, users can become
``friends'' by requesting and accepting friendship proposals.  Each user has
a wall on which all his or her friends can post messages directed to the
wall's owner.  According to \texttt{MySpace} policy, each user can also
see his or her friends' walls, in addition to its own wall.  
From the  automatically retrieved dataset of publicly available messages between
users, which we described in Ref.\ I, the 
 social network is reconstructed. The users are represented
as \emph{nodes}, directing messages to friends is represented by
\emph{directed links}, where the number of messages sent in the
particular direction determines the \emph{weight} $W_{ij}$ of the
directed links from $i$ to $j$.

In Ref.\ I we have analysed
topology of the network for two different time depths, two and three
months. Here we shortly summarize some topology features which are
relevant for this work.
Regarding the topology by omitting the weight of the link, we observe
a different distribution for the out-degree, i.e., the number of friends a
user sends messages to, and the in-degree, i.e.,  the number of friends a
user receives messages from. The out-degree distribution follows a power
law with a scaling exponent of 2.5, that does not change much if the time
window is extended from two to three months. On the other hand, the
in-degree distribution changes, converging to a stretched exponential when the time
window is increased. Eventually, we found some disassortative mixing with
respect to in-degree (``who is linked to my neighbors who are writing to me?''), which remains stable if the time window is increased
\cite{we-MySpace11}. These networks are very sparse with the link density $\rho \equiv L/N_U(N_U-1)$ in the range $2.8-3.3 \times10^{-5}$ and small reciprocity of links $r\equiv (L^{\leftrightarrow}/L - \rho)/(1-\rho)$ close to 0.021 and practically independent on the time depth.

\begin{figure}
\centering
\begin{tabular}{cc}
\resizebox{28.8pc}{!}{\includegraphics{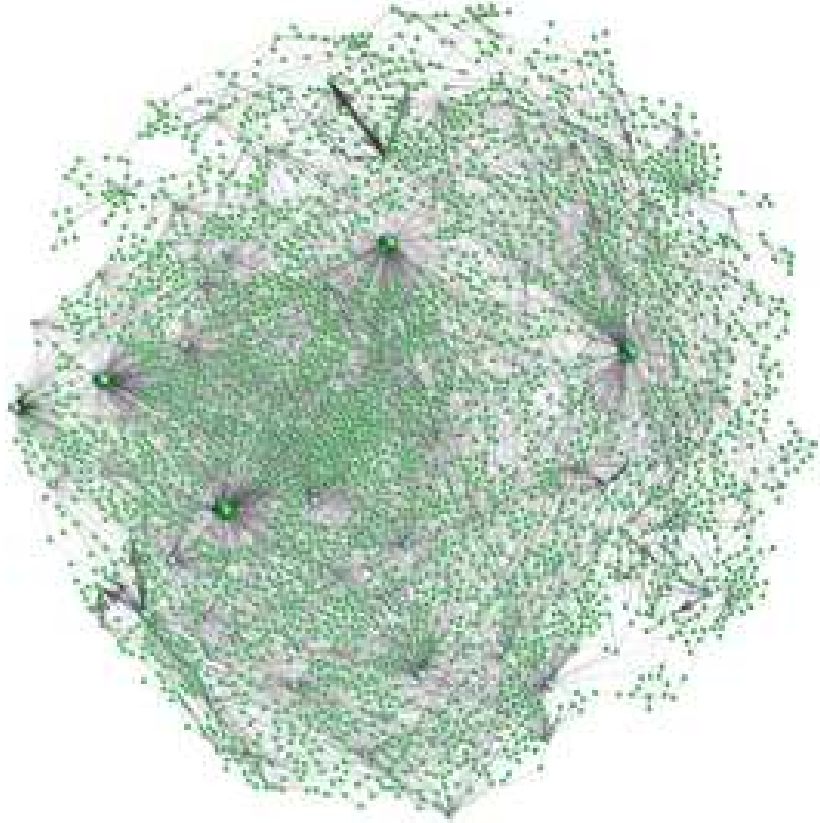}}
\end{tabular}
\caption{View of the directed network \texttt{Net3321} on which the simulations are performed. It contains $N_U=3321$ nodes and weighted links, based on the dialogs of two months depth among the users in \texttt{MySpace}, data from \cite{we-MySpace11}. }
\label{fig-Net3321}
\end{figure}

Here we use a part of the network termed \texttt{Net3321} extracted from the dataset  of two-months time depth. The network is reduced and compacted, the users that have sent and received less
than 5 messages within two months have been excluded. Hence the reduced  network  contains $N=3321$  nodes organized in four user communities \cite{we-MySpace11}. Its average link density $\rho=7.18\times 10^{-4}$ and reciprocity $r=0.118$. 
Graphical view of  \texttt{Net3321} is shown in Fig.\ \ref{fig-Net3321}, where the widths of the directed links correspond to the number of messages retrieved from that dataset.

\subsection{Properties of driving signal $\{p(t)\}$ and user delay times $P(\Delta t)$\label{sec-parameters}}
\begin{figure}[!h]
\centering
\begin{tabular}{cc} 
\resizebox{24.8pc}{!}{\includegraphics{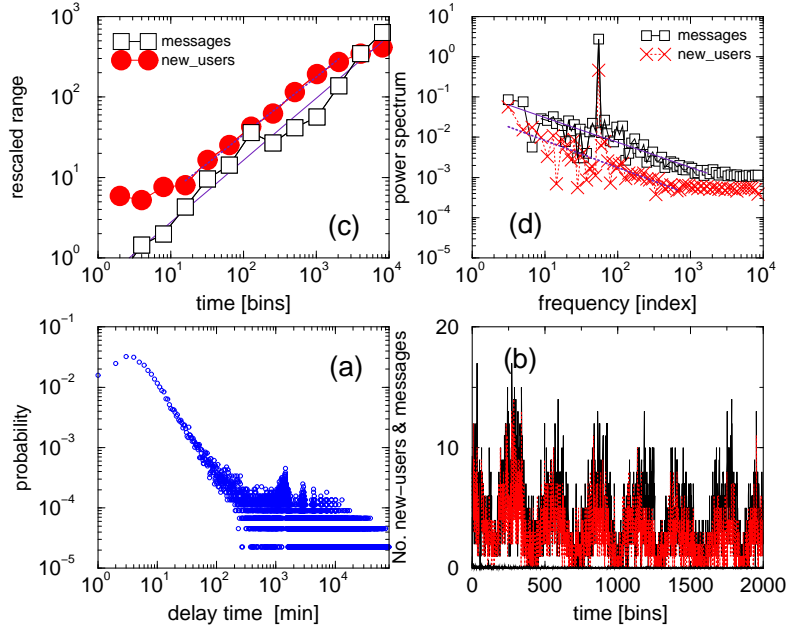}}\\
\end{tabular}
\caption{Activity patterns of the \texttt{MySpace} dataset for the time
window of two months:
(a) Distribution of the
  inter-activity time of users in the original resolution; (b) Time
  series of the number of new users $p(t)$ (red) and number of messages
  $N_c(t)$ (black line) per 5-minutes time bins. 
(c) Persistence and (d) Power spectrum of the time series of
  new users (red) and number of messages (black).  See text for more
  details. Both analyses are smoothed by logarithmic binning.}
\label{fig-fromdata}
\end{figure}

The \texttt{MySpace} dataset further allows us to obtain other empirical
features about the user activity, which are later used to calibrate
the model.  As it will be clear in Sec.\ \ref{sec-model}, we use the delay times distribution $P(\Delta t)$ and the time series of new user arrivals $p(t)$ as the input for the model. In this Section we extract and analyse these quantities from the empirical dataset. The results are shown in Fig.\ \ref{fig-fromdata}.

In \cite{we-MySpace11}, this distribution $P(\Delta t)$ was determined from the two datasets. It is fitted with a power-law  with a small exponent for the delay times longer than one day. 
Fig. \ref{fig-fromdata}a shows the interactivity-time
distribution $P(\Delta t)$ of users, where $\Delta t$ refers to the time
between two consecutive actions (delay) of the same user, with the original time resolution of $t_{\mathrm{bin}}=1 \mathrm{min}$.   In this paper we are just using the histogram of the interactivity-time as an input for our computer simulations as explained in Sect. \ref{sec-model}

The second important empirical value is the number of new users $p(t)$
entering the network per time unit $t$, which is shown in
Fig. \ref{fig-fromdata}b (red line) with a time resolution of $t_{bin}=5\
\mathrm{min}$. The time series indicates large fluctuations, but also
some periodicity. In the same figure we also show the total number of
messages $N_c(t)$ per time unit (black line) with the same time
resolution, which reveals that $N_{c}(t)$, which is the driven signal,
tightly follows $p(t)$, which is the driving signal. In conclusion,
communication in \texttt{MySpace} seems to be driven to a large extent by
new users entering the network for the first time (with respect to the beginning of the time window).

To further verify this correlated behavior, we analysed the Hurst
exponent $H$ and the power spectrum $S(\nu)$ of both $p(t)$ and $N_c(t)$ time series, as shown in Fig. \ref{fig-fromdata}c,d. The Hurst exponent in the two time series  measures the persistence of
the community dynamics, i.e. how consistent the dynamics is (i) in
attracting new users and (ii) in generating new messages, respectively.
Note that the time series in Fig.\ \ref{fig-fromdata}b are stationary. The exponent 
$H$ is calculated as follows: For both time series $p(t), N_{c}(t)\to x(t)$ with
total length $T$,  the series is divided into $n$ boxes of equal time span. Then the maximal range of the fluctuations $D_Y(n)$ of the cumulative time series $Y_n\equiv \sum_{t=1}^n(x(t)-<x>)$ is found and compared with its standard deviation $\sigma (n)$ on the same interval $n$. Plotting the rescaled fluctuations range  $D_Y(n)/\sigma (n)$ against $n$, the exponent is extracted from the power-law dependence as $D_Y(n)/\sigma (n) \sim n^H$, for $n=1,2, \cdots T$.  The plots in Fig.\ \ref{fig-fromdata}c show the rescaled fluctuations for the time series of new users and the number of messages with exponents $H_p=0.89\pm 0.06$ and $H_{N_c}=0.77\pm 0.04$, respectively.

The power spectrum $S(\nu)$ is the transformation of the time series
$p(t)$, $N_{c}(t)$ in the frequency space $\nu$, as shown in
Fig. \ref{fig-fromdata}d. One notice the remarkable peak in the
spectrum which results from the daily pattern of users online regularly.
Further, we find that, in the regions of low frequencies $\nu$, both
spectra are characterized by a power-law decay $S(\nu)\sim \nu^{-\phi}$,
with similar exponents $\phi_{p} =0.67\pm 0.11$ and
$\phi_{N_{c}}=0.61\pm0.09$, respectively. However, the power-law range is almost
one order of magnitude smaller for the driving signal $p(t)$ than for the
driven signal $N_{c}(t)$.  This suggests that the system builds
correlated behavior on the shorter time scales.  This is also seen in the
differences between the persistence ranges for small $n$ in
Fig.\ref{fig-fromdata}c. But we emphasize that $p(t)$ cannot be treated
as a proxy of $N_{c}(t)$ because of the difference. Instead, there is a
variable that links the two, which is related to the writing activity. As
we will show later, this link is given by the emotions of the users
expressed in terms of valence and arousal.
In our model we will use the driving signal $p(t)$, i.e. the number of
new users entering the social network, as an input variable.

\subsection{Extraction of emotional content from message texts}
\label{class}

So far, we have described the topology of the social network and the
activity patters of the users. This provides the basis for the most
important question, namely how users interact and influence each other,
which is the focus of our paper. For this, we have to analyse the
messages exchanged between users, in particular their emotional content.
Specifically, we quantify emotions with respect to two dimensions,
arousal and valence \cite{russell1980}. The latter indicates the
pleasure associated with the emotion (positive, negative, neutral), the
former the level activity that it induces.

In order to extract the emotional content from the messages, we use
sentiment analysis by applying a standard procedure introduced in
\cite{Dodds2009}. It uses the ANEW dataset, a {\it lexicon of human
  ratings of valence and arousal} with about 1000 words
\cite{Bradley1999}, for which the emotional charge, or valence, $v_{i}$
and the arousal $a_{i}$ was determined. An algorithm then calculates the
frequency $f_{i}$ of such classified words in a given text message, to
compute the valence and arousal of the text sequence as
\begin{equation}
  v_{\rm text} = \frac{\sum_{i=1}^{n}{  v_i f_i}}{\sum_{i=1}^{n}{ f_i}} 
 \;; \quad a_{\rm text} =  \frac{\sum_{i=1}^{n}{a_i f_i}}{\sum_{i=1}^{n}{ f_i}} 
\label{eq-ANEW}
\end{equation}
For the first time,  here we apply this method to the \texttt{MySpace}
dataset. Due to the limited size of the ANEW lexicon, the method should
be preferably used on long texts because of statistical reasons. To
overcome this limitation, we extracted the stem, or root form, of all
words in the analyzed text. The stem contains most of the semantic
information of a word (and thus its emotional content), and allows us to
match similar words rather than exact matches, which eventually improves
the statistics. To extract the stem, we used Porter's Stemming algorithm
\cite{VanRijsbergen1980}, a technique that applies inverse generalized
rules of linguistic deflection, mapping deflected words to the same
stem. For example, the stem of the words ``lovely'' and ``loving'' by
that method is ``love'', which matches the corresponding word in the ANEW
lexicon.  This way, the sentiment analysis covers a larger portion of the
text and allows to calculate the emotional values for more than 60\% of
the messages in the dataset.

\begin{figure}[h]
\centering
\includegraphics[width=0.40\textwidth]{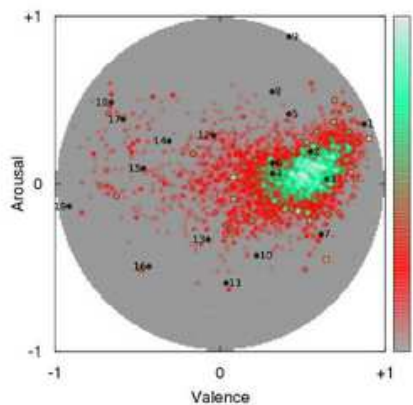}
\caption{Circumplex map of the emotions extracted from the
  \texttt{MySpace} dataset. The frequency is indicated by the color
  scale from 0 to 50, and the binning is done in a grid of 200x200
  bins.  Markers indicate different emotions: 1-``delighted'',
  2-``amused'', 3-``interested'', 4-``expectant'', 5-``convinced'',
  6-``passionate'', 7-``hopeful'', 8-``feeling superior'',
  9-``astonished'', 10-``longing'', 11-``pensive'', all on positive
  valence side, and 12-``impatient'', 13-``worried'',
  14-``suspicious'', 15-``distrustful'', 16-``ashamed'',
  17-``frustrated'', 18-``disgusted'', 19-``miserable'', on the
  negative valence side. }
\label{fig-circumplex}
\end{figure}

The results of the analysis of emotional expression of the \texttt{MySpace} dataset
are presented in Fig.  \ref{fig-circumplex}.  To be compatible with
Russell's circumplex model \cite{russell1980}, we rescaled the output of
Eqs.\ (\ref{eq-ANEW}) to the range $[-1,1]$ and adopted the standard polar
diagram for the quantitative representation of emotions by using the
following transformation:
\begin{equation}
  \label{eq:circumplex}
 x' = x \sqrt{1 - \frac{y^{2}}{2}} \;; \quad
 y' = y \sqrt{1 - \frac{x^{2}}{2}}
\end{equation}
Different points in this diagram are associated with different emotions.
For comparison, the markers $1-19$ indicate examples of emotions which
are known in psychology \cite{scherer2005}.

\begin{figure}[!h]
\centering
\includegraphics[width=0.32\textwidth]{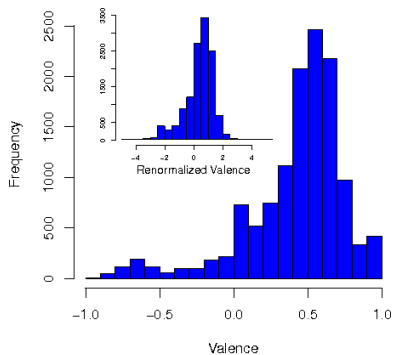}
\caption{Distribution of estimated valence of messages with at least
  one word from the ANEW dataset. Inset: distribution of the
  renormalized valence taking into account the amount of ANEW words in
  the message and the statistics of human expression on the Internet
  from \cite{garcia2011}. Both histograms show a clear bias towards positive
  expression, even above the natural bias of human expression.}
\label{fig-valdist}
\end{figure}

The distribution of valence values for the messages of our dataset is
highly biased towards positive values, as the largest density in
Fig. \ref{fig-circumplex} is above 0. Fig. \ref{fig-valdist} shows the
distribution of valence for messages that contain at least one word from
the ANEW dataset. We notice that the mode is at 0.5. On the other hand,
we have shown \cite{garcia2011} that English written text is naturally
biased towards positive emotions, with a mean $\mu_{v} = 0.31$ and a
standard deviation $\sigma_{v}=0.47$ for the valence.  This needs to be
taken into account to interpret Fig. \ref{fig-valdist}, so we renormalise
each valence value using $v' = {(v - \mu_{v})}/{\sigma_{v}/\sqrt{w}}$,
where $v$ is the valence from Eq.\ (\ref{eq-ANEW}) and $w$ is the amount of
ANEW words in the message. The renormalized valence distribution is shown
in the inset of Fig. \ref{fig-valdist}. We find that, despite this
renormalization, there is still a large bias towards positive emotions in
the messages of \texttt{MySpace}.

On the other hand, the distribution of the expressed arousal (vertical
axis of Fig. \ref{fig-circumplex}) is quite concentrated around values
close to 0, i.e. \texttt{MySpace} messages rarely contain words
expressing strong arousal. This finding is in line with previous survey
studies \cite{paltoglouIEEE2011} in which arousal from written texts
showed a small variance. 

\section{Model of emotional agents on \texttt{MySpace} network\label{sec-model}}

\subsection{Emotional dynamics of agents\label{sec-rules}}
Our agent-based model of emotional influence between \texttt{MySpace}
users builds on the empirical findings about (i) their interaction
network, (ii) their temporal activity patterns, (iii) the entry rate of
new users, and (iv) the emotionality of their messages. However, we do
not get from the data the way emotional messages influence the activity
and the emotional state of other users. Thus, in our model we provide
hypotheses about this feedback which are tested against the aggregated
outcome.

In specifying the model, we follow the agent-based framework of
emotional influence outlined in \cite{schweitzer2010}, which was
already applied to product reviews \cite{garcia2011b}, chat rooms
\cite{garas2011}, and blog discussion \cite{MMBT_ABM2011}. In this
framework, the emotional state of agent $i$ is described by two
variables, valence $v_i(t)\in [-1,1]$, and arousal $a_i(t)\in [0,1]$
which each follow a stochastic dynamics. 
In previous work \cite{garcia2011b,garas2011} stochasticity was modeled simply by an additive stochastic force.  In the present model 
we assume, similarly to the model of agents on blogs \cite{MMBT_ABM2011}, that stochasticity may result from three sources: (i)
sampling from the empirical inter-activity time distribution $P(\Delta
t)$, Fig. \ref{fig-fromdata}a, (ii) sampling from the empirical rate
$p(t)$ at which new users enter the network, Fig. \ref{fig-fromdata}b,
(iii) a spontaneous reset of both valence and arousal to a predefined
value $(\tilde{v},\tilde{a})$ with a probability $p_0$. 
The latter captures uncertainty in determining the external influences on an agent's state and is treated as a tunable parameter, as explained below. The
value of this state resets can vary according to very different
origins, including influential events for the users or emotional
consequences of the design of \texttt{MySpace} as a website. Another
source of stochasticity is on choosing the receipient of the message,
as explained below.

We further introduce an internal binary variable $\theta_{i}$ which
describes if an agent is active ($\theta_{i}=1$) or not
($\theta_{i}=0$). The value of $\theta_{i}$ is set according to the
sampling from the inter-activity time distribution $P(\Delta t)$. If the
agent is not active, it only relaxes towards a neutral state
$(v,a)=(0,0)$ with decay rates $\gamma^{v}$, $\gamma^{a}$. If it is
active, however, the agent perceives messages displayed on its own or on
friend's walls which affect its valence and arousal.  We note that this
definition of ``active'' is somewhat different from previous work where
activity always implied writing a message. Here, writing is considered a
subsequent activity which only happens with a certain probability
dependent on the agent's arousal and on the global activity level which
is proxied by the entry rate $p(t)$ as described below.

\begin{figure}[h]
\centering
\includegraphics[width=0.48\textwidth]{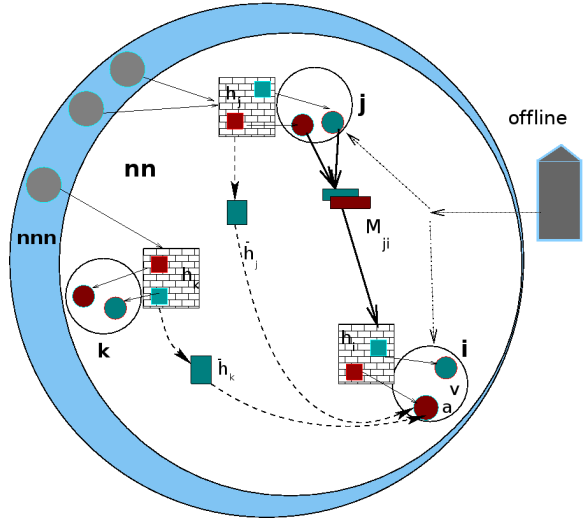}
\caption{Schematic presentation of three levels in our model which influence arousal $a$ and valence $v$ of the agent $i$: (1) first neighborhood $nn$ (agents $j$ and $k$) can write messages on the agent's $i$ wall, message stream, e.g.,  $M_{ji}$ from $j$ carries  arousal and valance of the agent $j$; aggregated aging messages on 
$i$'s wall, represented by the field $h_i$ with its arousal  and valence components (shown by small squares) directly contribute to the arousal and valence of $i$; Moreover, by viewing the walls of neighbors $j$ and $k$, and according to their valence similarity  with agent's $i$ current valence, the fields $\overline{h_j}$ and $\overline{h_k}$ are built and their arousal components contribute to $i$'s arousal. (2) next-nearest neighbors $nnn$ have message streams to first neighbor's $j$ and $k$ walls, thus  level (1) can be affected. (3) events in the off-line world can occasionally  have direct influence to  each agent by reseting their  arousal and valance to certain value $(\tilde{a},\tilde{v})$.  }
\label{fig-modelschema}
\end{figure}

To model how agents perceiving messages are affected by these, there are
three levels of aggregated information in our model: (i) aggregation of
messages on the agent's wall which shall be captured by an information
field $h_i$, (ii) aggregation of messages perceived on the friends'
walls, captured by an information field $\overline{h}_i$, (iii)
aggregation of messages on all walls, i.e. a mean-field information
$h_{\mathrm{mf}}$ that that captures a kind of ``atmosphere'' of the
community.  Fig. \ref{fig-modelschema} provides a simplified  overview of the
different contributions to the information field and how they affect the
emotional state of two agents $i$ and $j$. Because each of the messages is
 carrying  a valence and arousal value ( according to Eq. (\ref{eq-ANEW})in analogy to real data),
the information field $h$ also has a valence and arousal component
$h^{v}$, $h^{a}$ which results from the respective
aggregation. Specifically, different from previous modeling assumptions
\cite{schweitzer2010, garcia2011b}, we assume here
that the agent's valence is only affected by the valence information, while the 
arousal and activity (e.g. in choosing conversation
partners)  are primarily affected by the arousal information, but also include the contribution of the valence fields, as explained below.

Using discrete time, the dynamics of the emotional state of agent $i$ is
decribed by:
\begin{align}
\label{eq-stateDecay}
  v_{i}(t+1)=& (1-\gamma_{v})v_{i}(t) + \delta_{\theta_{i},1}\mathcal{F}_{v}(t) \\ \nonumber
  a_{i}(t+1)=& (1-\gamma_{a})a_{i}(t) + \delta_{\theta_{i},1}\mathcal{F}_{a}(t)
\end{align}
Here $\delta_{\theta_{i},1}$ is the Kronecker delta which is one only if
the agent is in an active state and zero otherwise. The nonlinear
 functions $\mathcal{F}$ capture how valence and arousal are
affected by the information fields $h$. An analytical discussion for these
was provided in \cite{schweitzer2010}. Here, we use the following
assumtions:
\begin{align}
  \mathcal{F}_{v}(t) =&
[(1-q)h^{v}_{i}(t)+qh^v_{mf}(t)]
\times [c_1+c_2(v_{i}(t)-v^{3}_{i}(t))][1-\vert v_i(t)\vert ]
  \nonumber \\
  \mathcal{F}_{a}(t) = & 
\{(1-q)[\epsilon h_{i}^{a}(t) +(1-\epsilon)\bar{h}^a_i(t)]+qh^a_{mf}(t)\}
\times [1-a_i(t)]
\label{eq-myspacearousal}
\end{align}
Each of these functions consist of a term that depends on the information
fields  and a second term that depends on the arousal or the valence,
respectively. For the latter, nonlinear assumptions are made in
accordance with \cite{schweitzer2010}. The term $[1-\vert x\vert]$ is added to
confine both variables in the prescribed range of the phase space. 
The small parameter $q\in[0,1]$ adjusts the fraction of the 
influence of the mean-field information, $h_{\mathrm{mf}}(t)$, whereas
the small parameter $\epsilon\in[0,1]$ adjusts the influence of the
information of the friends' walls, $\bar{h}_{i}$, in relation to the own
wall, $h_{i}$. The latter makes a difference between the valence and the
arousal dynamics. We assume that the arousal dynamics of an agent depends
on the activity on the walls of its friends, captured by
$\bar{h}^{a}_{i}$. We neglect this influence in the valence, because the
level of pleasure should rather depend on the messages directed to $i$
and not to its neighbors.

In order to close the model, we now have to define the dynamics of the
different components of the information fields. For our investigation of
the social network we assume that the mean-field information can be
neglected, i.e. $q=0$, whereas for other applications such as blogs dynamics \cite{MMBT_ABM2011} some mean-fields influence ($q=0.4$) is needed, while in the product reviews \cite{garcia2011b} or chat room conversation \cite{garas2011} the
mean-field plays the main role, i.e. $q=1$.  Moreover, in the case of social networks, \textit{the main
contribution of the information field comes from the agent's individual
wall}, thus we set $\epsilon=0.9$. We define the sequence of all messages
from agent $j$ to agent $i$ as $M_{ji}$. So the wall of an agent captured
by $h_{i}$ contains the message sequences from all its friends. The
influence of this information, however, decays in time with a rate $\gamma^{h}$. For the
valence and arousal component of $h$, we assume the following dynamics
(where $z$ stands either for valence $v$ or arousal $a$): 
\begin{equation}
h^{z}_i(t)= 
\frac{\sum_j \sum_{m \in M_{ji}}\theta(t,t_m)z_j(t_m)W_{ji}e^{-\gamma^h(t^{lm}_{ji}-t_m)}}{\sum_j \sum_{m \in M_{ji}} \theta(t,t_m) W_{ji}e^{-\gamma^h(t^{lm}_{ji}-t_m)}}e^{-\gamma^h (t-t^{lm}_{ji})} 
\label{eq-hiaw}
\end{equation}
$h_{i}$ is generated by steams of messages $M_{ij}$ from the neighbor nodes, where for each messages $m$ creation time $t_{m}$ is traced and the
Heavyside function $\Theta[x]$ ensures that its influence does not
start before. The emotional content of the message is composed by the
values of valence or arousal $z_{j}(t_{m})$ of the neighboring agents $j$ at
time $t_{m}$. Its influence further depends on the weight $W_{ji}$ of the
directed link which is determined empirically (see
Sect. \ref{sec-empirical}). The exponential decay terms result from our
assumption of the decay of the information influence with rate
$\gamma^{h}$, where $t^{\mathrm{lm}}_{i}$ is the time of the last message
on the wall of agent $i$. The denominator of eq. (\ref{eq-hiaw}) plays
the role of a normalization to keep the field values properly bounded.

In addition to the individual field component $h_{i}$, the influence of
the friends' walls aggregated in $\bar{h}_{i}$ has to be specified. As
explained above, there is only the arousal related component
$\bar{h}^{a}_{i}$ for which we assume: 
\begin{equation}
\bar{h}_i^a(t)=
\frac{\sum_j W_{ij}h_j^a(t)(1+h_j^v(t)v_i(t))}{\sum_j W_{ij} (1+h_j^v(t)v_i(t))}\ ,
\label{eq-secondterm}
\end{equation}
Firstly, $\bar{h}^{a}_{i}$ is composed of the average over the weighted
arousal fields on the friends' walls at time $t$. The weights $W_{ij}$,
however, are modified by a term that takes into account the similarity
between the valence $v_{i}$ of the agent $i$ and the valence of its friends
walls captured by the valence field $h_{j}^{v}$. This is due to the fact that agents can
only see the message walls of their friends, but not directly observe
their valence.  Information thus becomes the more important the more it is in
line with the emotions of an agent. There is also a psychological
argument for this assumption: users often search for information
reinforcing their emotional state \cite{Bradley2009}. Further,
there is a technical argument: to cope with information overload
\cite{walter2008} most social networking sites filter
information such that content presented to the user is in line with its
previous writing.  We note that Eq.\ (\ref{eq-secondterm})
captures the influence of every ``friend-of-a-friend'', because they are
able to post messages on the walls of the friends of an agent, i.e. there
is a second-nearest-neighbor influence that cannot be perceived in the
same way in off-line social interactions, but is very characteristic for
online social interaction. 
Our model captures this important feature of the online communication dynamics. 

The last part of specifying the model regards the writing activity of the
agent which is subsequent to perceiving messages. If an agent is in the state
$\theta_{i}=1$, it write a message with a probability $\omega_{i}$ that
increases with its current arousal $a_{i}$. The proportionality, as mentioned
before, depends on the global activity level proxied by $p(t)$ and a
strength parameter $a_{0}$, hence $\omega_{i}(t)=\delta_{\theta_{i},1}
a_{0}p(t)a_{i}(t)$. The emotional content of the messages is given by the
emotional state of the agent in terms of $v_{i}$ and $a_{i}$ at the time of writing.  Next, the
recipient of the message has to be determined. Instead of a uniform
random choice, friend $j$ of an agent will be chosen with a probability
$s_{j}(t)$ that depends on (i) the aggregated information $h_{ji}$
generated by $j$ on the wall of $i$ (whereas $W_{ji}$ represents the
strength of the social link estimated by the total number of messages
generated by $j$ on $i$'s wall), and (ii) the importance of the wall of
friend $j$ to agent $i$. The rational behind this assumption is the
following: when a user writes a message to someone else, this can be part
of an ongoing conversation or initiate a conversation. The former is
reflected in the first term and the latter in the second term of the
following equation:
\begin{equation}
s_j(t)={\cal{A}}[\beta \frac{W_{ij} h^a_{ji}(t)}{\sum_k W_{ik}} +
(1-\beta) \frac{W_{ij} h^a_j(t) (1+h^v_j(t)v_i(t))}{\sum_k W_{ik}
(1+h^v_k(t)v_i(t))}],
\label{eq-probpj}
\end{equation}
${\cal{A}}$ is the inverse normalization constant and $\beta$ is a parameter
weighting between these two influences of the neighbors versus the own
experience. For simplicity, we later choose $\beta$ equal to $\epsilon$,
which weighted the information in the neighbors' walls against the own
wall. The aggregated information  $h_{ji}$ along the $j\to i$ link is assumed to depend only on the arousal component of the sender $j$ in the following way: 
\begin{equation}
h^{a}_{ji}(t)=\frac{\sum_{m \in M_{ji}} \theta(t,t_m)
a_j(t_m) e^{-\gamma^h(t^{lm}_{ji}-t_m)}}
{\sum_{m \in M_{ji}} \theta(t,t_m) e^{-\gamma^h(t^{lm}_{ji}-t_m)}}
e^{-\gamma^h (t-t^{lm}_{ji})},
\label{eq-hij}
\end{equation}
Similar to Eq.\ (\ref{eq-hiaw}), this aggregates a set $M_{ji}$ of messages coming from the agent $j$ on the wall of $i$.

\subsection{Simulation setup and numerical implementation}
\label{sec:setup}

In the previous section, we have specified the emotional dynamics of
agents with respect to valence and arousal, the activity of agents in
perceiving and creating messages, and communication between agents in
terms of the dynamics of the communication field consisting of different
components. To numerically investigate this agent-based model, we need to
specify the setup in terms of the network, the initial conditions of
agents, and the parameters. 

As stated in Sect.\ \ref{sec-network}, the network used for our
simulations is taken from the empirical data of \texttt{MySpace}, 
and consists of $N=3321$ agents.  Their weighted links $W_{ij}$ are calculated from the two-months time window. 
The initial values of the agents' emotional states are chosen
uniformly at random from the intervals $a_i\in [0,1]$, $v_i\in
[-1,1]$. Also, their activity pattern is drawn from the empricial
inter-activity time distribution $P(\Delta t)$, shown in
Fig. \ref{fig-fromdata}a. While we are sampling from a distribution
observed at resolution of $t_{\mathrm{res}}=1$ min, the time scale $t$ of our simulations
is fixed by the driving signal $p(t)$ with  $t_{bin}=5$min time bin, i.e., each time step in the simulations corresponds to one time bin of real time. 
 Hence, in case of sampling $\Delta t < 1$, the
event happens at the current time step. Note that this includes five possible values of the delay time, to which different probability is assigned according to the  high-resolution distribution in Fig.\ \ref{fig-fromdata}a.  
The global activity level on the
social network is captured by the empirical signal $p(t)$,
Fig. \ref{fig-fromdata}b, as explained both in Sect. \ref{sec-empirical} and
when outlining the model.

Apart from the empirical distribution function $P(\Delta t)$ and the time series $p(t)$ that we use as input for the simulations, the values for the control parameters introduced in the model are specified as given in
Table\ \ref{tab-parameters1}. We can distinguish between parameters
influencing the emotional state of the agents (``internal'',
$\gamma^{a}$, $\gamma^{v}$), parameters controlling the communications
between agents ($\gamma^{h}$, $q$, $\epsilon$) and parameters that
control the activity of agents ($p_0$, $a_{0}$). We recall that $p_0$
measures the probability at which the emotional state of the agent is
determined externally to a value $(\tilde{v},\tilde{a})$. For this,
two different setups are discussed in the results: (i)
$(\tilde{v},\tilde{a})$ are chosen at random from a uniform
distribution, and (ii) $(\tilde{v},\tilde{a})$ are fixed to three sets
of values, each representing a target emotion (a) ``ashamed'', (b)
``enthusiastic'', (c) ``astonished''.  The first case corresponds to the null
hypothesis of a completely unknown external influence that might come
from any kind of influence to the agent's emotional state. The second
case of fixed values of $(\tilde{v},\tilde{a})$ will allow us to test
the collective effects of large scale events or mass media, by
assuming their influence in the emotional state of the users of
\texttt{MySpace}. Furthermore, design decisions of the website might
externally drive the emotions of its users to particular values, and
we can simulate the collective effects of such decisions in the
context of our model.

\begin{table}[!h]
  \caption{Values of the control parameters and the input functions used in the simulations. }
\label{tab-parameters1}
\centering
\begin{tabular}{|llll|c|}
\hline
 Internal parameters&Decay rates& Influence & Driving & Global functions \\
\hline
$c_1=0.5$& $\gamma^a=\gamma^v=0.05$& $q=0$& $p_0=0.01$& $P(\Delta t)$ \\
$c_2=0.5$& $\gamma^h=0.01$& $\epsilon=0.9$& $a_0=0.01$& $p(t)$\\
\hline
\end{tabular}
\end{table}

\begin{algorithm}
\caption{Program Flow: Emotional Agents on Social Network  \texttt{MySpace}}
\label{alg-ABMmyspace}
\begin{algorithmic}[1]
\STATE \textbf{INPUT:} Network $\{W_{ij}\}$; Parameters $\epsilon$, $q$, $c_1, c_2$, $a_0$, $\gamma^v$,
$\gamma^a$, $\gamma^h$, $p_0$, $p(t)$, $t_{max}$;
Distribution $P(\Delta t)$; 
\STATE Set initial conditions for $a_i(0)$ and $v_i(0)$;
\FORALL {$1 \le i \le N$}
\STATE {set user delay  $\Delta t(i)$ to next random from $P(\Delta t)$ distribution;}
\ENDFOR
\FORALL {$t < t_{max}$}
 \STATE compute  mean fields $h_{mf}^a(t)$, $h_{mf}^v(t)$; 
\FORALL {$1 \le i \le N$}
 \STATE compute all local fields
 $h^a_{i}(t)$, $h^v_{i}(t)$,  $h^a_{ij}(t)$;  
\ENDFOR
 \FORALL {$1 \le i \le N$}
  \STATE with probability $p_0$ set $v_i(t)$ and $a_i(t)$ to uniform random value/or another predetermined value;
 \ENDFOR
 \FORALL {$1 \le i \le N$}
  \IF {$\Delta t(i)>t_{bin}$}
   \STATE nonactive user: relax user valence $v_i(t)$ and arousal $a_i(t)$ with the rate $\gamma_v$,$\gamma_a$;
\STATE decrease their $\Delta t$;
  \ELSE
   \STATE active user: update  arousal $a_i(t)$ and valence $v_i(t)$  using Eqs. (\ref{eq-stateDecay}) and (\ref{eq-myspacearousal});
    \STATE with probability $a_0 p(t) a_i(t)$ create message $m$ to target selected using distribution given with Eq.\ (\ref{eq-probpj});
   \STATE set $\Delta t(i)$ to next random from $P(\Delta t)$ distribution;
  \ENDIF
 \ENDFOR

\STATE Sampling the data of interest;
\ENDFOR
\STATE {\bf END}
\end{algorithmic}
\end{algorithm}
The agent-based model is numerically implemented in C++ and the pseudo code is given in Algorithm\ \ref{alg-ABMmyspace}. 
We note that the emotional states of the agents are
updated in parallel. At each time step, first all field components for
the wall of each agent are updated. Second, with probability $p_0$ the
emotional state of each agent is set to a value
$(\tilde{v},\tilde{a})$. Third, the activity state of each agent is
determined with respect to the inter-activity time distribution $P(\Delta
t)$. Specifically, each agent has an internal counter related to its
activity, which is reduced by one in every time step. Agents whose
current counter is lower than $t_{bin}$ become active.  For these, a new
value of the counter is sampled from the inter-activity distribution and
their emotional state is updated. With probability $\omega_{i}$, a
message is created and the recipient is chosen with probability $s_{j}$.

\section{Results of Computer Simulations}
\label{sec-simulations}

\subsection{Individual trajectories}
\label{sec:traj}

\begin{figure}[!h]
\begin{tabular}{cc} 
\resizebox{16.0pc}{!}{\includegraphics{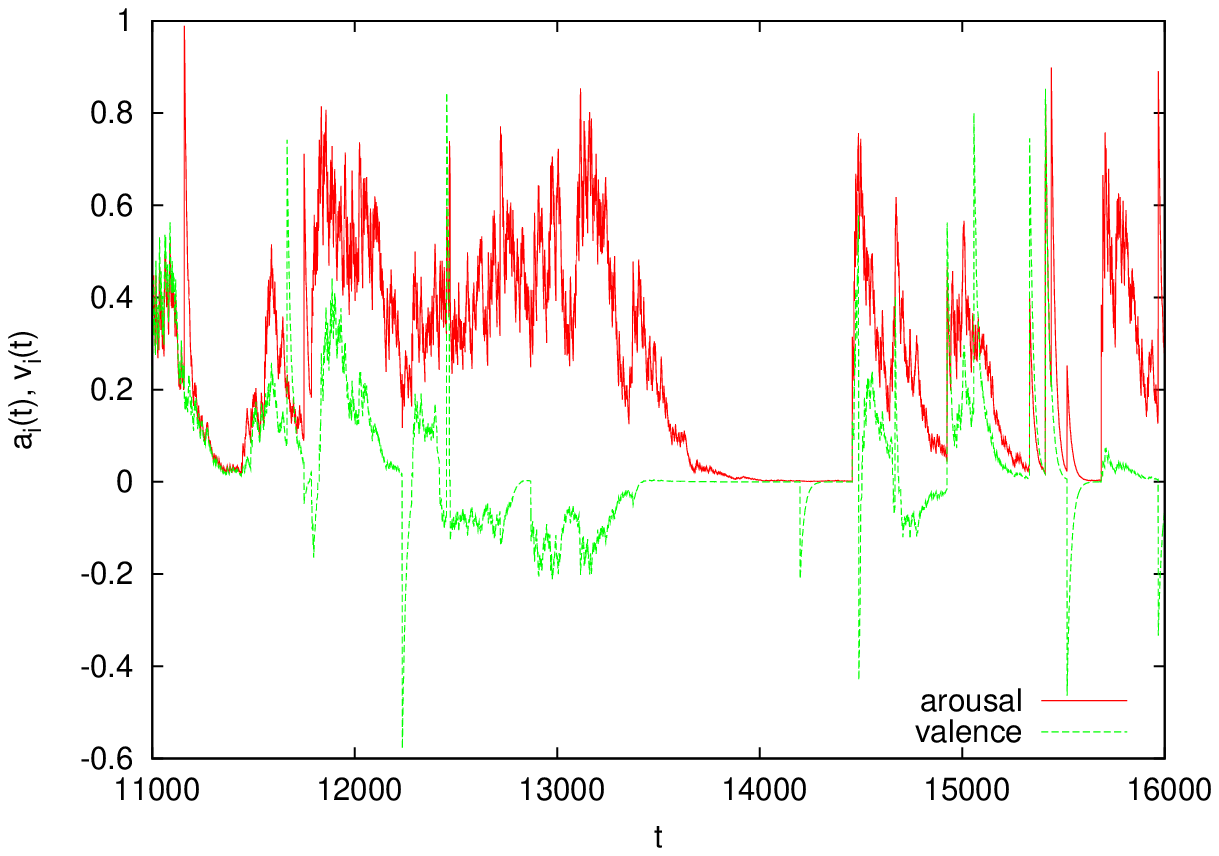}}&
\resizebox{16.0pc}{!}{\includegraphics{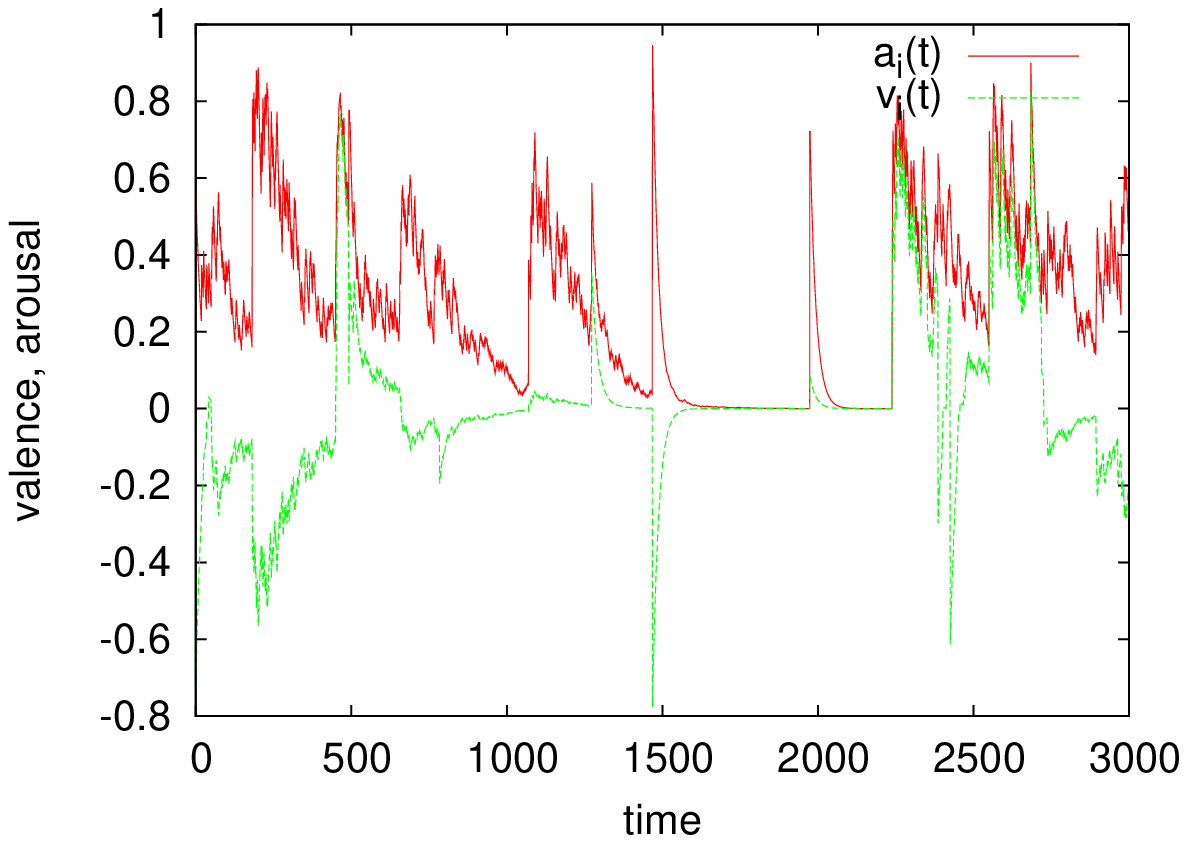}} \\ 
\end{tabular}
\caption{Time series of the valence (green) and the arousal (red) of
  two active agents in a simulation of our \texttt{MySpace} model}
\label{fig-singleagent}
\end{figure}

In this paper, we have provided an agent-based model in which agents
influence each other by exchanging emotional messages. The model captures
essential features of user interaction in \texttt{MySpace}. Before we
discuss the aggregated output, we provide two examples of the individual
dynamics of agents in terms of their emotional variables valence $v_i(t)$
and arousal $a_i(t)$. The simulations shown in Fig. \ref{fig-singleagent}
demonstrate that states with high arousal can be built up due to the
interaction of an agent with its neighborhood or, less often, by a
single large input from its environment. Repeated activities on short
time intervals occur if an agent is 'caught' by an active neighborhood
over long periods of time. One notices in these time series the peaks in
both arousal and valence when the agent was influenced by a message from
a neighbor, or when its emotional state was externally reset to
$(\tilde{v}, \tilde{a})$. In the absence of any action both valence and
arousal exponentially decayed towards zero. Further, one notices that
agent's valence can be influenced either toward positive or negative
values, without specific preferences.

\subsection{Valence distribution}
\label{sec:valence}

We now measure the applicability, or even the success, of our model by
comparing its aggregated output with empirical results from
\texttt{MySpace}. Some of the empirical findings are already used as an
input for the simulations, as explained in Sects.\ \ref{sec-empirical},
\ref{sec:setup}. However, the activity patterns and emotional response of
agents do not trivially follow from a driving signal $p(t)$ and a given
network topology. Instead, only with the right assumptions about the
agent's emotional interaction, we are able to reproduce the stylized
facts as explained below. For the following simulation results, we sample
the external influence $(\tilde{v}, \tilde{a})$ from a uniform
distribution, to then explore later the effects of these parameters in the
collective behavior of the model.

\begin{figure}[!h]
\centering
\begin{tabular}{cc}
\resizebox{24.8pc}{!}{\includegraphics{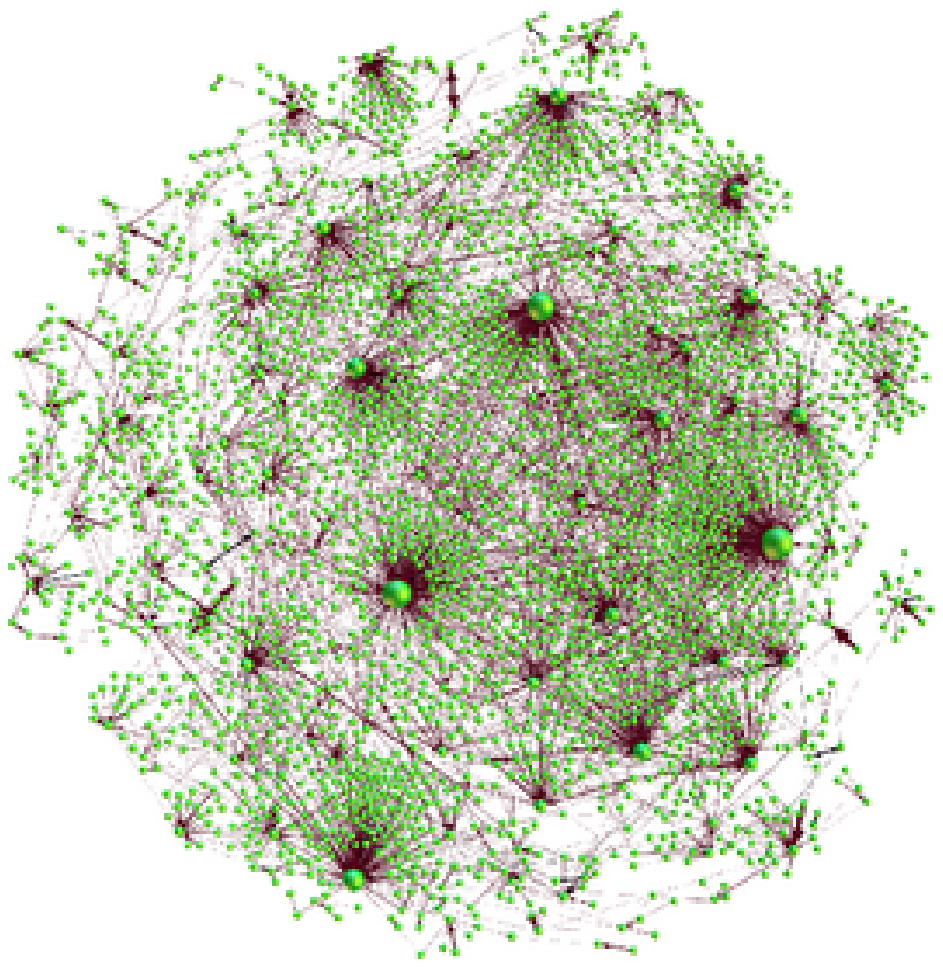}}\\
\resizebox{24.pc}{!}{\includegraphics{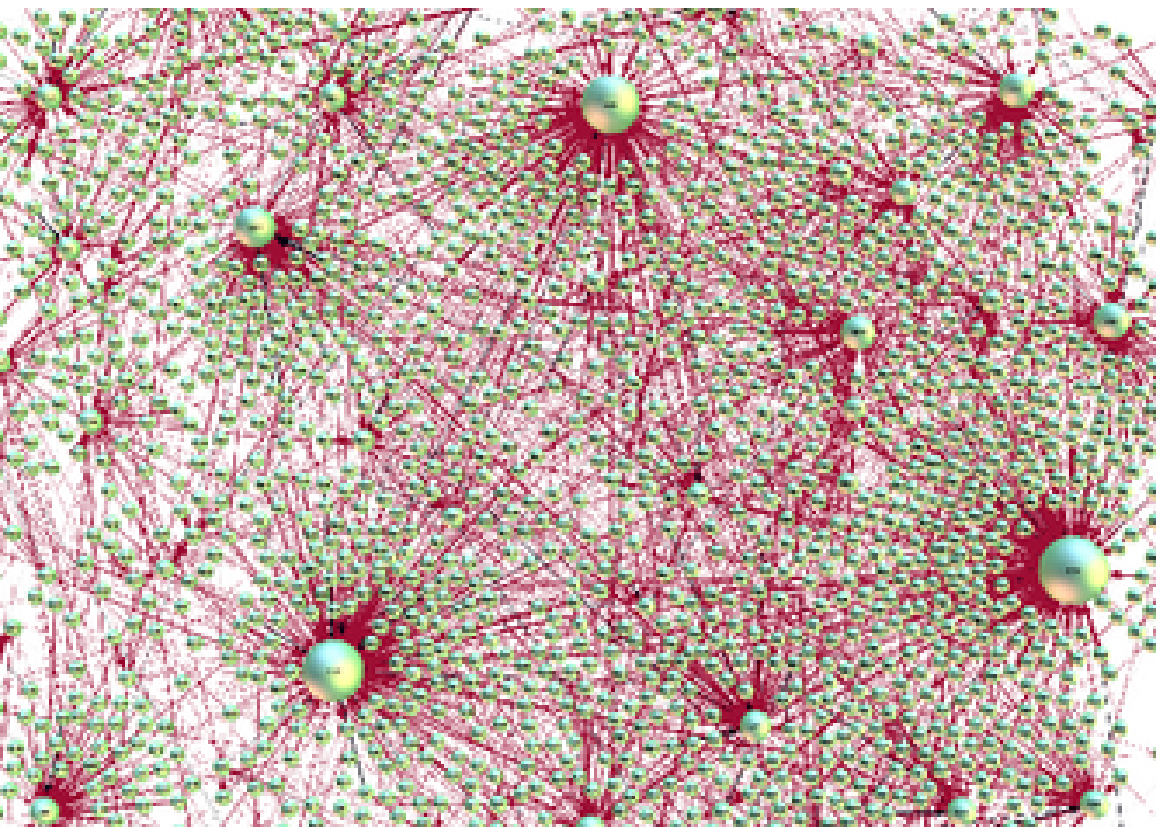}}\\
\end{tabular}
\caption{An example of simulated dialogs between the agents as nodes
  on the part of \texttt{MySpace} network \texttt{Net3321} (top) and zoomed part of it (bottom).  The agents received the external influence
  with $(\tilde{v}, \tilde{a})$ set to the ``enthusiastic'' emotion state.  
Two nodes are connected by a directed edge that
  symbolizes the communication (at least one message) from one agent
  to another during the simulation time. Edge width is proportional to
  the number of messages sent along the the link, and the edge colors
  are chosen according to the average valence expressed in these
  messages: negative (black), positive (red).  
}
\label{fig-simnetwork}
\end{figure} 

We start by explaining similarities in the network of emotional
interactions. The network topology is given empirically and does not
change through the simulations. What changes instead is the activity of
the (fixed number of) agents and their emotional states and \textit{which links on that network are used} to transmit emotion. 
Fig. \ref{fig-simnetwork} shows, for the topology of the \texttt{Net3321} 
the aggregated emotional content of the
messages exchanged, at the end of the simulation. The coloring pattern could be compared to the empirical pattern in paper I \cite{we-MySpace11} Fig.\ 7. 
Specifically,  the links are
directed, their width is given by the amount of messages sent through
that link during the simulation time, and their color represents the average valences of the messages sent through that link. It is obvious that there is a strong bias towards generating messages with positive emotions (indicated by red color). 

\begin{figure}[!h]
\centering
  \includegraphics[width=0.4\textwidth]{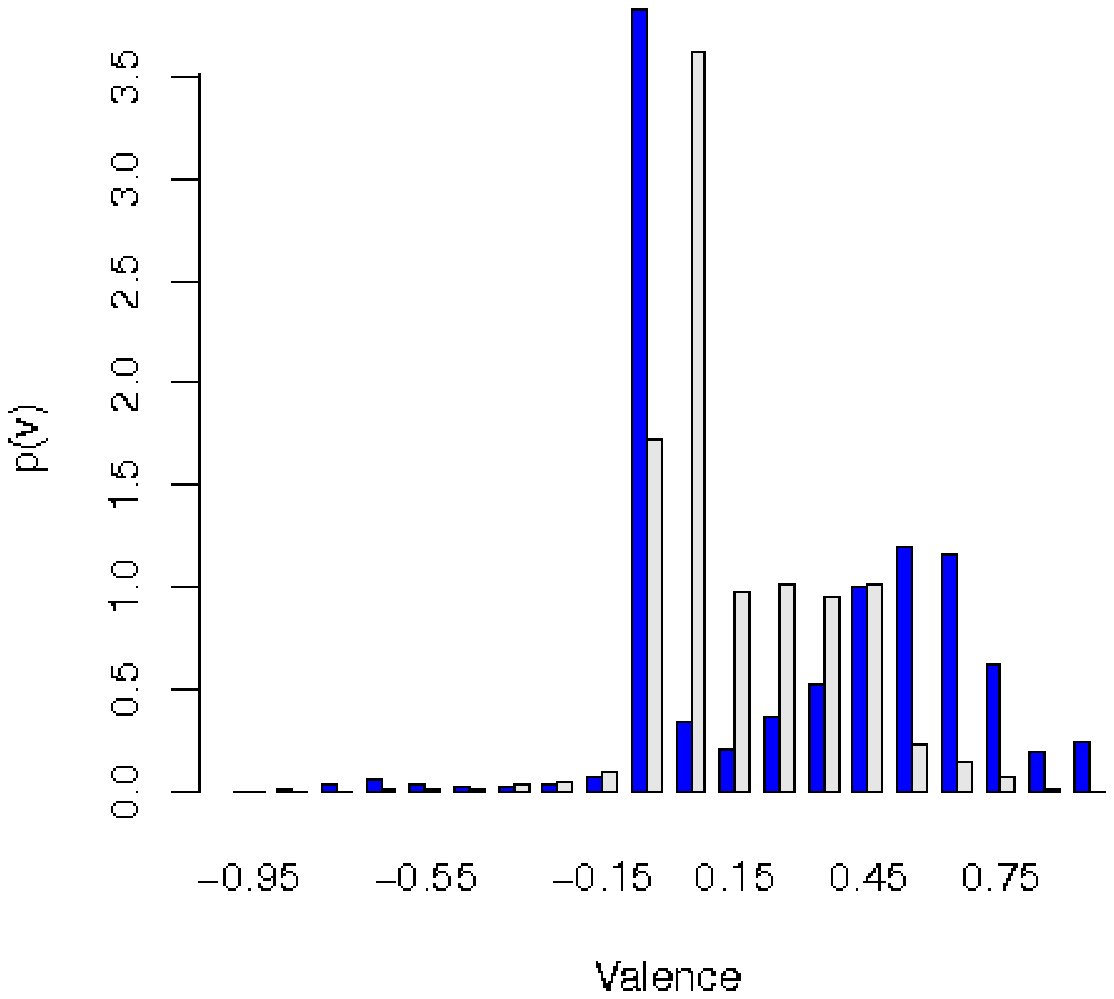} 
\caption{Distribution of the valence of \texttt{MySpace} messages (dark),
  including the ones without any ANEW term in the bin at $v=0$. In
  grey, the distribution of expressed valences in the simulation
  depicted in Fig.\ \ref{fig-simnetwork}. Both distributions have a
  large bin of nonemotional expression close to zero, and the rest of the
  distribution shows a strong bias towards positive values. }
\label{fig-valDistSim}
\end{figure}

We can compare the valence distribution of the messages generated by the
agents with the one of the users. Fig. \ref{fig-valDistSim} shows both
the empirical and the simulated valence distribution aggregated over
time. The empirical distribution is equivalent to the one shown already
in Fig.\ \ref{fig-valdist}, except for the peak at $v=0$ which contains
all messages that did not contain any word from the ANEW dataset used for
classification. We notice that both distributions have an obvious bias
toward positive valence, which is stronger in the empirical data than in
the simulations.  

But instead of arguing about quantitative differences, we emphasize that
our model is indeed able to reproduce this bias as the result of
emotional interactions. Without any interaction, the agent's valence
would relax towards zero because of the decay factor in
Eqs.\ (\ref{eq-stateDecay}). However, the social interaction with other
agents through the fields $h$ on the network generates the positive bias 
in agreement with the empirical findings.  This also supports previous work that
argues about the social origin of the positive bias \cite{garcia2011,Rime2009}.

\subsection{Activity patterns of agents}
\label{sec:actsim}
\begin{figure}[!h]
\centering
\begin{tabular}{cc} 
\resizebox{20.4pc}{!}{\includegraphics{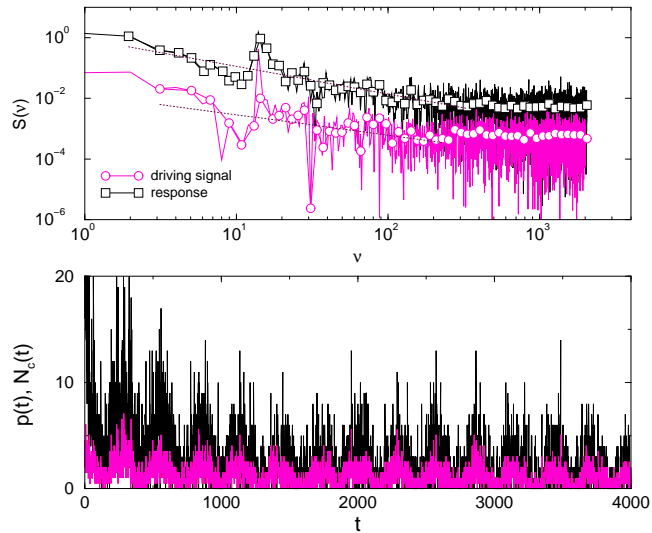}} \\
\end{tabular}
\caption{
Simulated time-series of the number of messages $N_c(t)$ in black and the
    driving signal $p(t)$ in pink of the emotional agents on the network \texttt{Net3321} (lower panel) and its power spectrum (upper panel). The straight lines indicate the slopes of the correlated part of the power-spectrum.
}
\label{fig-myspaceABMtimeseries}
\end{figure}

The empirical findings about user activities have been discussed in
Sect.\ \ref{sec-empirical}. We recall the skew distribution of the
inter-event time for users, $P(\Delta t)$, Fig. \ref{fig-fromdata}(a),
and the number of new users $p(t)$ and total messages $N_{c}(t)$ over
time, Fig.\ \ref{fig-fromdata}(b), which showed a strong daily periodicity. 
 As already explained in Sect.\ \ref{sec-empirical}, in the simulations we use both $P(\Delta t)$ and $p(t)$ as inputs for our computer simulations, which
implies that we need to reproduce the correct behavior of $N_{c}(t)$
using our model. 
The simulated results are shown in
Fig. \ref{fig-myspaceABMtimeseries} both in terms of the time series and
the power spectrum $S(\nu)\sim \nu^{-\phi}$, which can  be compared to the ones in  Fig.\ \ref{fig-fromdata}(b,d).  
While $ \phi_{p} = 0.67 \pm 0.11$ for the driving signal was given, we obtain $\phi_{N_{c}} \approx 0.91 \pm 0.08$  
for the driven signal, which deviates from the empirical value
$\phi_{N_{c}} \approx 0.65 \pm 0.12 $ given in Sect.\ \ref{sec-empirical}.
First, we note that we get indeed a time series for $N_{c}(t)$ which
shows the long-term correlation expected. However, the crossover between
the uncorrelated and the correlated events occur at a larger time scale
in the simulations, i.e., cascades on small time scales are not captured
by the simulations. On the other hand, the behavior on long time scales
is well reproduced, i.e. our model is able to reproduce the emergence of
long range correlations in the emotional expressions, which means we see
the emergence of collective emotions.

\begin{figure}[!h]
\centering
\begin{tabular}{cc} 
\resizebox{20.8pc}{!}{\includegraphics{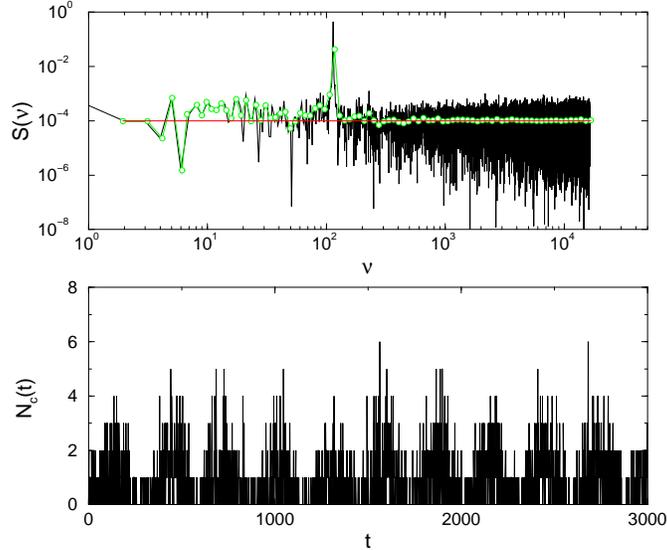}} \\ 
\end{tabular}
\caption{(bottom) Time series and (top) power-spectrum of the number of
  messages $N_c(t)$ in a simulation as response to an artificial
  driving signal composed of white noise with circadian cycles.}
\label{fig-addPSpectra-2}
\end{figure}

Let us now discuss the importance of the spectral properties of the
driving signal $p(t)$. Would we get a long-range correlation in the total
number of messages produced also with a non-correlated driving signal?
In order to test this, we performed simulations in which the driving
signal is composed of white noise with a circadian component of daily
periodicity. The average value of the white noise is set to the mean of
the empirical time series, $\langle p(t) \rangle$, but there are no long
range correlations, in contrast to the original signal $p(t)$.  Keeping all other
parameters unchanged, we simulate the network response to this driving signal. 
Contrary to the results in Fig.\ \ref{fig-myspaceABMtimeseries}, 
now $N_c(t)$ does not show any correlations, as
displayed in Fig.\ \ref{fig-addPSpectra-2}. 
I.e., apart from the daily periodicity,  the power spectrum retains the characteristics of a white noise.
Occurrence of the circadian cycle of daily activity is not enough to create the
long range correlations present in the real data. 
We comment on this in again in the conclusions. 

\subsection{External influences in community behavior}
\label{sec:ext}

To understand the influence of external events on the emergence of
collective emotions, our simulations consider externally triggered
resets of the emotional state of agents to a value $(\tilde{v},
\tilde{a})$.  As we explained in Sec. \ref{sec:setup}, $(\tilde{v},
\tilde{a})$ can be either fixed or sampled from a uniform
distribution. While the latter case is covered by the simulation
results explained in Sect. \ref{sec:actsim}, we now use fixed values
for $(\tilde{v}, \tilde{a})$ which capture three different external
events, for example in mass media: (a) ``astonished'' (v=0.4, a=0.88),
(b) ``ashamed'' (v=-0.44, a=-0.5), (c) ``enthusiastic'' (v=0.5,
a=0.32).  In the psychology literature \cite{scherer2005}, 
``astonished'' and ``ashamed'' are assumed to have an influence on the
social interaction and emotional communication.  In our quantitative
model, these two emotional states are in the opposite parts of the
circumplex map (see Fig. \ref{fig-circumplex}): ``astonished'' is a
positive emotion with high arousal, while ``ashamed'' is a negative
emotion with low arousal.

\begin{figure}[!h]
\centering
\begin{tabular}{cc}
\resizebox{24.8pc}{!}{\includegraphics{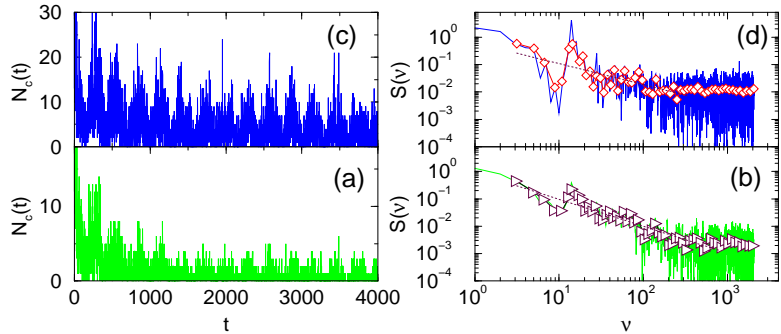}}\\
\end{tabular}
\caption{(a) Simulated time series $N_c(t)$ and (b) their power
  spectra $S(\nu)$ in the case of fixed external influence
  $(\tilde{v},\tilde{a})$ set to the point corresponding to
  ``ashamed'' (Fig. \ref{fig-circumplex}). (c) and (d) same for  the
  values corresponding to the state ``astonished''.  In both cases the
  driving signal $p(t)$ was used.}
\label{fig-addPSpectra-AA}
\end{figure} 
We want to test how fixed values of $(\tilde{v}, \tilde{a})$ influence
the cascade of emotions on the social network. The simulated time series
and power spectra of $N_c(t)$ shown in Fig. \ref{fig-addPSpectra-AA}
correspond to either ``astonished'' or ``ashamed''.  The power spectra of
$N_c(t)$ in both cases show the same shape of $1/\nu^\phi$ with the 
exponents close to $0.76\pm0.11$ in the case  ``astonished'', while 
$0.98\pm 0.05$ approaching flicker noise, in the case  ``ashamed''. 
This values are higher but not significantly
different to the one found in the case of uniformly distributed
$(\tilde{v}, \tilde{a})$, shown in Fig. \ref{fig-myspaceABMtimeseries}.
But correlations extend for a larger range in the case of ``ashamed''
(also noticeable in the lower value of $S(\nu)$ at high $\nu$), even
though it is a low-arousal, negative emotion.  However, the level of
activity in the time series is higher for ``astonished'', as expected.

 The collective emotional response that is observed in the correlated
 time series, is built on the actions of individual emotional agents on
 the network. The evolution of the emotional state of an agent like the
 ones shown in Fig. \ref{fig-singleagent} can be seen as a trajectory in
 a circle like the one in Fig. \ref{fig-circumplex} by applying the
 transformation of eqn. \ref{eq:circumplex}. We can explore the
 collective effects of the different types of external influences by
 looking the emotional states ``visited'' by the agents on the
 circumplex \cite{ahn2010}.

\begin{figure}[!h]
\centering
\begin{tabular}{cc}
\resizebox{24.0pc}{!}{\includegraphics{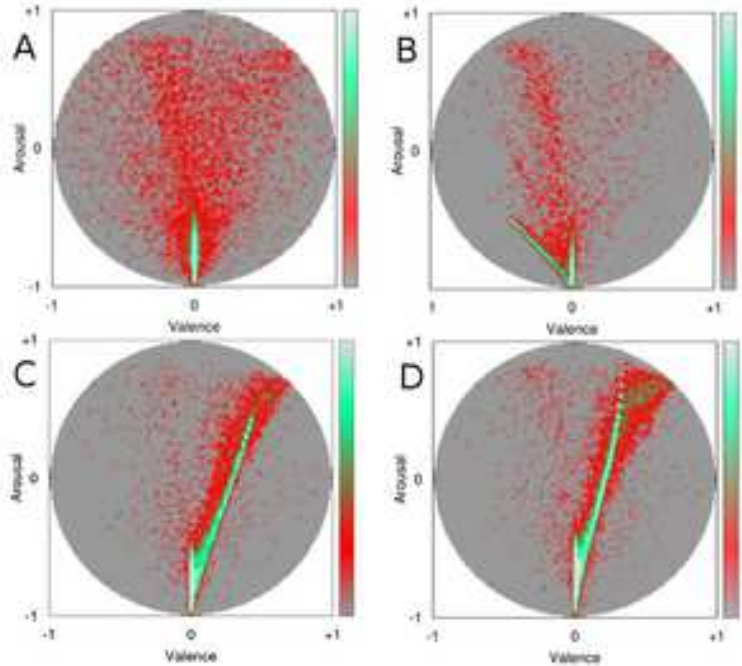}} \\
\end{tabular}
\caption{ Histograms of visited states in the circumplex for
   simulations with external influences sampled from a uniform
   distribution (a) and in fixed to ``ashamed'' (b), ``enthusiastic''
   (c) and ``astonished'' (d). Arousal is rescaled to the interval
   $[-1,1]$. 
}
\label{fig-myspaceABMcircumplex}
\end{figure}

 In the color plots of Fig. \ref{fig-myspaceABMcircumplex} we show the
 histograms of the different emotional states that were visited by the
 agents in our simulations. More precisely, we plot the emotion states
of each agent $i$ at every moment when the agent was active,
 i.e. $\Theta_i=1$.  The level of activity among the agents varies a
 lot, depending on their location in the network and the set of events
 occurring in the agents neighborhood. Consequently an agent's
 contribution to this emotion histograms will vary according to their
 activity in a given simulation.  The patterns shown in
 Fig. \ref{fig-myspaceABMcircumplex} represent situations with random
 and three specific  external influences: ``ashamed'', ``astonished'',
 and ``enthusiastic''. The histograms shown in the two upper panels of
 Fig. \ref{fig-myspaceABMcircumplex} show how high arousal states may
 arise starting from (a) uniformly distributed random input or (b) a low
 arousal negative valence state like ``ashamed''. In the lower panels,
 the external influences are set to two different high arousal positive
 states: ``astonished'' and ``enthusiastic''. The influence of these
 states spreads through the networks by means of agent interactions
 through the field, giving raise to the asymmetrical V-shape pattern.

\subsection{Propagation of emotions in the network}
\label{sec:flow}
The agent-based model also allows us to understand how cascades of
emotional influence propagate through the real social network of
\texttt{MySpace}. Precisely, will positive emotions propagate along the
same social links as negative ones?  Obviously, agents may exchange
positive or negative messages with different preferred neighbors. To what neighbor the message will be sent
depends on the agent's past interaction along the link, strength of the influence fields and the valence similarity with the wall of the recipient agent. 
The global problem of finding the pattern of most frequently used links on the entire network is appropriately captured by the \textit{maximum-flow spanning tree} of that network. On these trees each node is attached to the rest of the tree by its strongest link.
Again, strength of a link between two agents is determined as the total amount of messages sent along that link during the simulation time.

\begin{figure}[!]
\begin{tabular}{cc} 
{\Large{(a)}}\\
\resizebox{24.8pc}{!}{\includegraphics{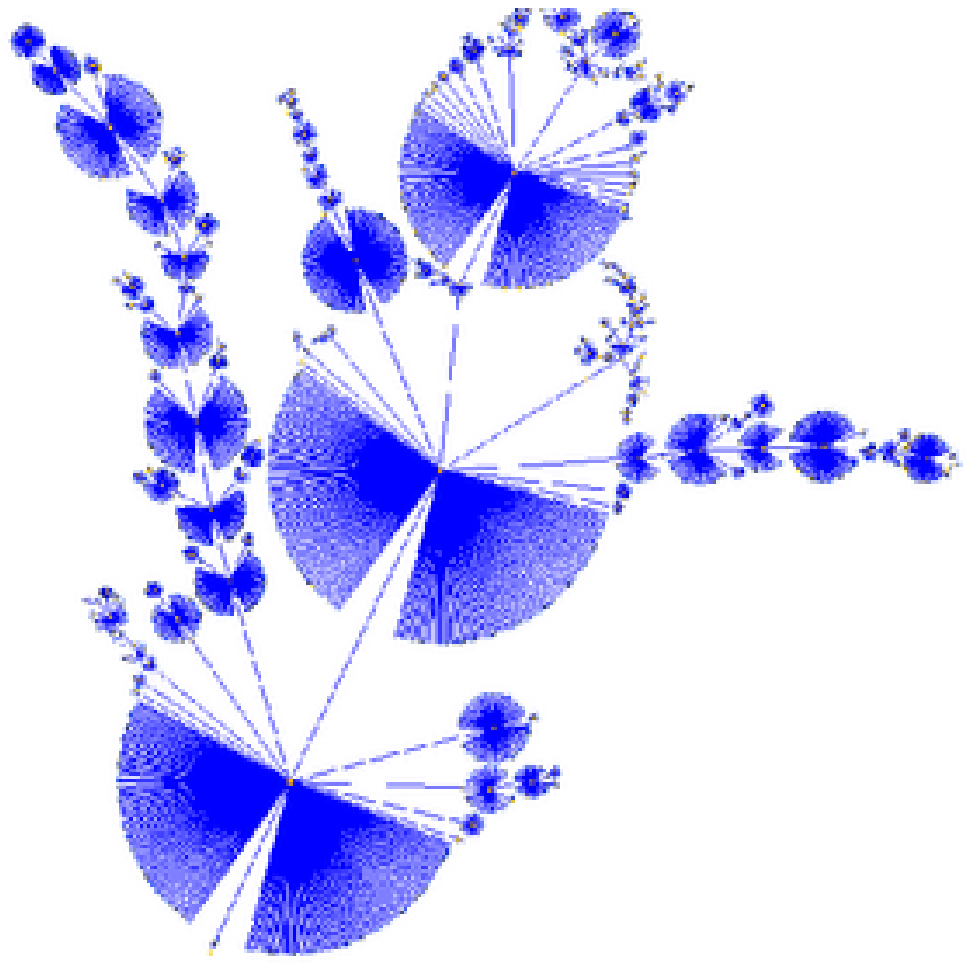}}\\
{\Large{(b)}}\\
\resizebox{24.8pc}{!}{\includegraphics{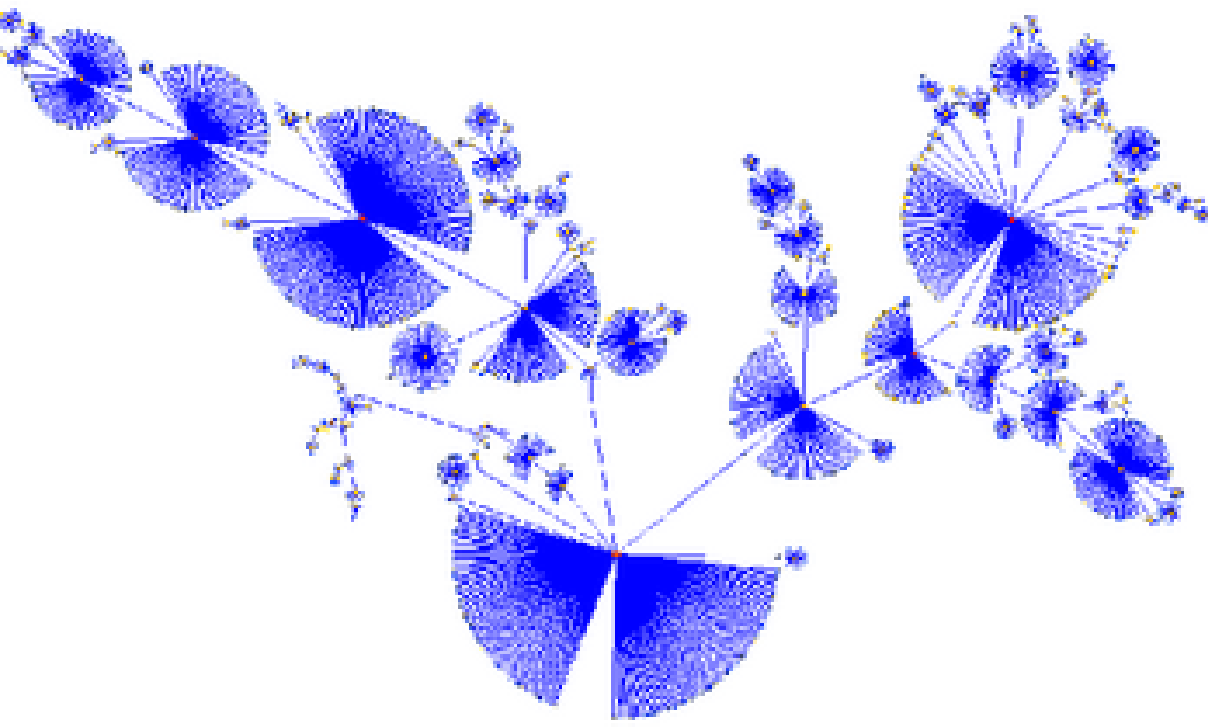}}\\
\end{tabular}
\caption{
Maximum-flow spanning trees for agent dynamics in our
  \texttt{MySpace} simulations with $(\tilde{v}, \tilde{a})$ fixed to
  the values correspoding to the state ``enthusiastic'' (top) and
  ``ashamed'' (bottom).}
\label{fig-myspaceABM_MFST}
\end{figure}

Fig. \ref{fig-myspaceABM_MFST} shows these strongest links between agents
for two simulations with different external influence of positive
(``enthusiastic'') and negative (``ashamed'') external events. For comparison, the time series from the same simulation runs are shown in Fig.\ \ref{fig-addPSpectra-AA} and the patterns of visited areas in the phase space, in Fig.\ \ref{fig-myspaceABMcircumplex}b,d. 
Obviously, for the whole network of $N=3321$ nodes shown, these flow patters differ considerably. 

We first note that the pattern reflects the directedness of the links,
i.e. agent $i$ may have its strongest link to agent $j$ (in terms of
messages exchanged), but not vice versa. This reveals the
existence of strong  hubs in the social network, i.e. agents to which many
other agents have their strongest links.  Interestingly, in the case of
the positive emotion with high arousal, ``enthusiastic'', the large hubs occur 
along the central branch of the tree, and similarly, side branches contain smaller hubs of comparable size. 
While in the case of negative emotions with low
arousal, ``ashamed'', the tree splits in two major branches and the hubs appearing along each branch are of different size. 

\section{Discussion and Conclusions\label{sec-discussion}}

We have studied dynamics of the emotion-driven dialogs in online
social networks with a particular emphasis on the emergence of
collective behaviors of users.  For this purpose we first analyzed a
large dataset of the dialogs collected from a part of \texttt{MySpace}
network. The analyzed data have high temporal resolution and contain
full texts of dialogs exchanged among users at the connected part of
the network within two months period. This enabled us to extract the
emotional contents---arousal and valence---from the texts the dialogs
and to study temporal correlations in their occurrence.  To further
understand the dynamics of such emotional communications, we have
introduced the model of emotional agents on the network.  In the
model, the users are represented by the agents, whose arousal and
valence fluctuate in time being influenced by internal and external
inputs and "reactivity'' of the network.  In full analogy to the
empirical data, a high-resolution dynamics is maintained in the model
with each message handled separately. The rules of actions are
motivated by realistic situations in the online social networks and
some of the parameters governing the dynamics are inferred from the
same dataset of \texttt{MySpace} dialogs.  Several other parameters, for
instance decay time of emotion and network "reactivity'', which can
not be estimated from the available empirical data, have been kept
within theoretically plausible limits. In addition, within the model
we infer the {\it action-delay} and {\it circadian} cycles as
generated by the real-world processes of \texttt{MySpace} users, which
condition the pace of their actions and stepping into the virtual
world of the online social network, respectively. With this "native'' set of control
parameters, the simulation results enable us to derive several
conclusions, in particular regarding the emotion spreading processes
in \texttt{MySpace}, and potentials of the model for predicting user behaviors
in hypothetical (experimental) situations.

Our main conclusions are summarized as follows:
\begin{itemize}
\item {\it Temporal correlations of users activity in \texttt{MySpace}} occur
  on long-time scale and are accompanied by high arousal and
  predominantly positive emotions.
\item {\it Rhythms of users stepping from real-to-virtual world} carry
  certain important features of the communication processes in the
  online social networks.  Specifically, the temporal correlations in
  the online dynamics are built as a response to already correlated
  step-in processes. Otherwise, if not driven in a different way, the
  online social networks with their internal dynamics of the
  user-to-user contacts and restricted visibility of messages are not
  capable to generate correlations on large temporal scale.

\item {\t Patterns of emotion dynamics} in the virtual world of social
  networks are different for positive and for negative emotions. In
  the empirical data of MySpace the positive-valence emotions
  dominate. However, model simulations of spreading emotional states
  with different arousal--valence components and different social
  connotations, ``enthusiastic'' and ``ashamed'', for example, show
  different patterns in the phase space of the emotions involved as
  well as the social links used to spread the emotions on the network.

\item {\it High-arousal states} in the dynamics are built on small
  noisy input for all initial emotion states in our simulations, which
  is reminiscent of ``party''-like behavior of the agents.  In our
  model this is a consequence of  collective effects---repeated
  actions of an agent caught in the active network environment. 
\end{itemize}

Quantitative analysis of the simulated and the empirical data lead to
similar results, for instance, comparison of the correlations of the
emotional time series in cf. Figs. \ref{fig-fromdata} and \ref{fig-myspaceABMtimeseries},
and range of the expressed valences, in Fig. \ref{fig-valDistSim}.
This suggests that the model of emotional agents can reproduce the
stylized facts of the empirical data of \texttt{MySpace} dialogs, when the
parameters are appropriately chosen!  Moreover, within the model,
genesis of the emergent behaviors with particular contributions of each
user (agent) and its social connections--can be revealed! This makes
the predictive value of the model. It is more subtle, however, to
relate the predictions of the model which regard the individual
agent's emotional state and its fluctuations with the "feelings
change" observed in the psychology research of the online
communications.  In this respect one can recognize that a
characteristic area in the positive-valence high-arousal states
recurrently being visited by the agents, may reflect the positive
baselines of human valence and arousal found in \cite{Kuppens2010}.

Moreover, the emergent asymmetrical V-shape patterns of
Figs. \ref{fig-myspaceABMcircumplex} correspond with the patterns of
natural selective attention discovered in psycho-physiological studies
\cite{Bradley1999}. For evolutionary reasons, humans have two modes of
reaction to emotional content: appetitive and defensive
motivation. Both tendencies can be seen in our simulations, opening
the question of which one of them is predominant in the users'
internal emotions. This way our agent based model provides testable
hypotheses for psychological research, as the dynamics of the
emotional reaction of users of online communities might depend on the
emotional state of new members. Experimental setups similar to the
ones presented in \cite{Kuster2011} can test whether the
physiological reactions to new community members (arrivals) follow the
patterns predicted by our model.

In conclusion, the presented agent-based simulations give a new insight into emotion dynamics in online social networks. In the model, the interaction rules, closely related to \texttt{MySpace} social network site, take into account influence of the next-neighborhood on the agent’s state—--a salient feature of the online social networks, and the extended phase space where common emotions can be recognized. Hence, despite of its mathematical complexity, our model provides a “laboratory” for further experiments on the emotional agent’s behavior (e.g., under different driving conditions, varied external inputs, and changed values of the parameters) and for a comparative analysis of online and offline social networks.

AUTHORS CONTRIBUTION:  Designed research: BT;  Developed the program code, and executed simulations: M.\v S.; Provided the tool for extraction of emotional content: D.G.;  Contributed with the concept of emotional agents: F.S.;
Analysed the data and contributed in graphics: B.T., M.\v S., D.G.; Wrote the paper: B.T., F.S., D.G.

{\bf Acknowledgments:}  {The research leading to these results has
  received funding from the European Community's Seventh Framework
  Programme FP7-ICT-2008-3 under grant agreement n$^o$ 231323 and the project P-10044-3.  
 B.T.  thanks support from the national program P1-0044 of the research agency of the Republic of Slovenia and COST-MP0801 action. 
M.\v S. also thanks the national research projects ON171037 and III41011 of the Republic of Serbia.}

\end{document}